\documentclass[aps,pre,preprint,showpacs,amsmath,amsfonts,preprintnumbers]{revtex4}
\usepackage{epsfig}
\usepackage{yfonts}
\usepackage{amssymb}
\usepackage{graphicx}
\swabfamily
\newcommand{\Op}[1]{{{\mathrm{\hat{#1}}}}}

\begin{document}
\title{Algorithm for simulation of quantum many-body dynamics using dynamical coarse graining}
\author{M. Khasin and R. Kosloff}
\affiliation{Fritz Haber Research Center for Molecular Dynamics, 
Hebrew University of Jerusalem, Jerusalem 91904, Israel}
\date{\today }

\begin{abstract}

An algorithm for simulation of quantum many-body dynamics having  ${\mathfrak{su}}(2)$ spectrum-generating algebra is developed. The algorithm is based on the idea of dynamical coarse-graining.  The original unitary dynamics of the target observables $-$ the elements of the spectrum-generating algebra $-$ is simulated by a surrogate open-system dynamics, which can be interpreted as  weak measurement of the target observables, performed on the evolving system. The open-system state  can be represented by a mixture of pure states, localized in the phase-space. The localization reduces  the scaling of the computational resources with the Hilbert space dimension $n$ by factor $\frac{n^{3/2}}{\ln n}$ compared to conventional sparse-matrix methods.  The guidelines for the choice of parameters for the simulation are presented and the scaling of the computational resources with the Hilbert space dimension of the system is estimated. 
The algorithm is applied to the simulation  of the dynamics of systems of $2\cdot 10^4$ and $2\cdot 10^6$ cold atoms in the double-well trap, described by the two-sites Bose-Hubbard model.

\end{abstract}
\maketitle
\section{Introduction}

Simulation of quantum many-body dynamics is a challenging task. The computational resources of solving the Schr\"odinger equation increase with the size of the system's Hilbert space $n$, which is very large in a typical many-body \cite{kohn}. 
Computational methods based on diagonalization scale as the the cube of the Hilbert space dimension.
This led to the development of propagation techniques which scale as $O(n \ln n)$ (semi-linearly) with the Hilbert space dimension
for an elementary propagation time-step. The number of time-steps required for the simulation on a fixed time interval scales with the energy uncertainty of the evolving state, which adds generically at least another factor $n$ to the computational cost.  As a result, the overall scaling of the computational resources for a fixed time interval becomes $O(n^2 \ln n)$. To achieve such a scaling different sparse-matrix techniques are required such as FFT \cite{k56}
or linear scaling approaches in solid state \cite{Goedecker}. Currently such methods are able to address quantum dynamics in
Hilbert spaces up to dimension of $\sim 10^{8}$. Eventually due to the unfavorable scaling the semi-linear techniques become unpractical.

As a consequence, other strategies have been developed to meet the challenge of the many-body quantum simulations.
A  popular approach is based on extensions of a mean-field approximation and is restricted to systems with a limited  scope of quantum correlations. An example is the powerful multi configuration time dependent Hartree (MCDTH) algorithm, where part of the correlations are deliberately added back  \cite{meyer}.  As the quantum correlations develop in the course of the evolution the scaling of the method becomes prohibitive. 

The present paper describes the first implementation of an alternative approach to many-body quantum simulations \cite{khasin081}, which focuses on the dynamics of a small subset of observables. The motivation for the approach  is to find a method of quantum-dynamical simulations, scaling with the size of the subset of target observables but not with the Hilbert space dimension of the system. Simulations with such scaling properties are  defined  in the present work as  efficient.   The basic idea of the method proposed in  Ref. \cite{khasin081} is to simulate the unitary dynamics of a subset of observables  by a surrogate \textit{open-system} dynamics. The target observables in the proposed scheme are elements of the spectrum-generating algebra (SGA) of the quantum system \cite{bohm}  and the surrogate open-system dynamics corresponds to  weak measurement \cite{diosi06} of the observables performed on the evolving system. 

The present study explores  basic steps required to generate an operational algorithm and the scaling of the computational resources with the Hilbert space dimension  for simulation of quantum many-body dynamics having ${\mathfrak{su}}(2)$  SGA.
The expositions is carried out through the study of a specific example,
the simulation of the dynamics of $N$ cold atoms in a double-well trap \cite{leggett,schumm,gati,gati07,vardi00,trimborn}. The system is modeled by the two-site Bose-Hubbard model \cite{mahan} with the Hamiltonian:
\begin{eqnarray}
\Op H=-\Delta \sum_{i=1,2} (\Op a_{i+1}^{\dagger} \Op a_i+\Op a_{i}^{\dagger} \Op a_{i+1})+\frac{U}{2N}\sum_{i=1,2} (\Op a_{i}^{\dagger} \Op a_i)^2 \label{bhgenral},
\end{eqnarray}
where $\Op a_i$ ($\Op a_i^{\dagger}$) is the creation (annihilation) operator of a particle in the $i$-th well. $\Delta$ describes the hopping rate and $U$ scales the two-body interaction.
The transformation  to the  operators
\begin{eqnarray}
\Op J_x&=&\frac{1}{2}( \Op a_{1}^{\dagger} \Op a_2+\Op a_{2}^{\dagger} \Op a_{1} ) \nonumber \\
\Op J_y&=&\frac{1}{2 i}( \Op a_{1}^{\dagger} \Op a_2-\Op a_{2}^{\dagger} \Op a_{1} ) \label{suops} \\ 
\Op J_z&=&\frac{1}{2}( \Op a_{1}^{\dagger} \Op a_1-\Op a_{2}^{\dagger} \Op a_{2} ) \nonumber
\end{eqnarray}
leads to the following Lie-algebraic form of the Hamiltonian
\begin{eqnarray}
\Op H=-\omega \Op J_x+ \frac{U}{N} \Op J_z^2 \label{bhlie},
\end{eqnarray}
where $\omega=2\Delta$. The set of single-particle operators $\{\Op J_x, \Op J_y, \Op J_z \}$  is closed under the commutation relation and spans the ${\mathfrak{su}}(2)$ SGA  of the  system. Examples of  many-body quantum systems with   ${\mathfrak{su}}(2)$ SGA include other systems such as the Lipkin-Meshkov-Glick model of interacting fermions \cite{lmg}. The Hilbert space of the system of $N$ atoms 
 carries the $n=N+1$ dimensional irreducible representation \cite{Gilmore} of the  the algebra, corresponding to the pseudo-spin quantum number $j=N/2$. Observables
$\Op J_x , \Op J_y , \Op J_z$ are the target
of the dynamical simulation, representing coherence, current and population difference
between the wells.

The corresponding surrogate open-system dynamics is driven by the Liouville-Von Neumann equation of motion:
\begin{eqnarray}
\dot{{\Op \rho}}=-i\left[\Op H,{\Op \rho}\right]-\gamma \sum_i\left[\Op J_i\left[\Op J_i,{\Op \rho}\right]\right] \label{lvn}
\end{eqnarray}
The simulation of the open-system evolution  (\ref{lvn}) is performed by averaging over solutions of the \textit{stochastic nonlinear Schr\"odinger equation} (sNLSE) \cite{gisin84,diosi88,gisin92}. The sNLSE governs a single realization of  the process of weak measurement of the target observables (\ref{suops}) performed on the system with the Hamiltonian (\ref{bhlie}). Parameter $\gamma$ in Eq.(\ref{lvn}), corresponding to the strength of  measurement, is  decreased in the course of the simulation until the open dynamics (\ref{lvn}) of  the target  observables  converges to the original \textit{unitary} dynamics, driven by the Hamiltonian (\ref{bhlie}). It is shown that the convergence is achieved at the magnitude of $\gamma$ which is still sufficiently large to drastically reduce of the computational complexity of the sNLSE, compared to the original Schr\"odinger equation.
The reduction of the computational complexity is achieved by virtue of the localization of solutions of the sNLSE in the associated phase space, induced by the measurement  \cite{khasin081,khasin082}. On the level of the density matrix the localization leads to a coarse-graining of the phase-space representation of the state, erasing the fine structure but leaving unaffected the dynamics of the target observables.

Localization of the sNLSE solution leads to  compression of the size of the time-dependent computational basis of the simulation to  $M\ll n=N+1$ states. In the convergent simulation on the time-interval of order $\ln n$, $M \lesssim \ln n$. As a consequence, the  memory resources and the cost of an elementary time-step of the computation scale as $\ln^2 n$. The number of time-steps in the simulation  on the time-interval of order $\ln n$ is $O(\sqrt{n}\ln n)$  for a localized state evolution. Averaging over  realizations of the weak measurement adds a  factor of $\sim 10^{3}-10^4$. This  factor is independent on $n$.   Therefore, theoretically, the overall scaling of the computational resources   is $O(\sqrt{n}\ln^3 n)$. Simulations using conventional sparse-matrix techniques scale as $O(n^2\ln^2 n)$. Therefore, even though the algorithm is not efficient in the sense of our definition, the scaling of the computational resources is  better by the factor of $n^{3/2}(\ln n)^{-1}$ than in conventional sparse-matrix approaches.

An additional important feature of the algorithm is that  simulations of the sNLSE for individual measurement realizations are independent  and can be performed in parallel. This parallelism is uncommon in conventional  algorithms for quantum dynamical simulations.

Section \ref{sec:TheAlgorithm}(A) summarizes the main ideas of Refs.\cite{khasin081,khasin082} and  motivates the choice of  the computational basis for  simulation of the target observables. Section \ref{sec:TheAlgorithm}(B) develops  a scheme for updating the computational basis using a global unitary transformation generated by the SGA .

Section \ref{sec:ApplicationToTheTwoSiteBoseHubbardModel} focuses on the application of the algorithm to  dynamics of a large number $N$ of cold atoms in a double-well trap. The choice of parameters of the simulation and the scaling of the computational resources with the Hilbert space dimension  are analyzed and  numerical results of the simulation for $N=2\cdot 10^4$ and for $N=2\cdot 10^4$ atoms are presented.   Section \ref{sec:Conclusionsalgo} summarizes  the results.

\section{The algorithm}
\label{sec:TheAlgorithm}
\subsection{General structure}

The algorithm proposed in Ref.\cite{khasin081}, is based on the following sequence of steps:
\begin{enumerate}
\item{Construction of the time-dependent computational basis of  generalized coherent states (GCS), associated with the SGA of the system.}
\item{The choice of  elements of  SGA of the system as the target observables for the simulation. }
\item{Numerical solution of the sNLSE, corresponding to a single realization of weak measurement of the target observables, performed on the evolving system.}
\item{Averaging the expectation values of the target observables  over a large number of realizations.}
\item{Convergence of the open-system dynamics of the target observables to their original unitary evolution, achieved by increasing the size of the computational basis and decreasing the strength of the measurement.}
\end{enumerate}

\subsubsection{The choice of the time-dependent basis}
\label{sec:TheChoiceOfTheTimeDependentBasis}
As a guideline for the development of the algorithm  the following definition of  efficient simulation is adopted:

\textbf{Definition:}\textit{ The simulation of dynamics of a Lie-algebra  of observables of the system is efficient if the necessary memory and the CPU resources for the computation do not depend on the Hilbert space (irreducible) representation of the algebra.}

In short, the simulation is called efficient if it is based on group-theoretical calculations.
An example of  efficient simulation is solution of the Heisenberg equations of motion for a Lie-algebra $\mathfrak{g}$ of observables, when the Hamiltonian belongs to the algebra. In this case the simulation is reduced to numerical solution of $\dim\left\{\mathfrak{g}\right\}$ linear equations of motion for the expectation values of the elements of $\mathfrak{g}$. Another example is simulation of  quantum dynamics in the mean-field approximation \cite{kramer}, when the computational complexity is independent on the Hilbert space dimension asymptotically.

The starting point for the algorithm is the idea that  reduction of the computational complexity can be achieved by using  time-dependent basis of a small number $M$ of states.  Let $\psi_i(t)$ be the time-dependent computational basis. The state vector of the system can be represented in the time-dependent basis as 
\begin{eqnarray}
\left|\Psi(t)\right\rangle= \sum_{i=1}^M c_i(t)\left|\psi_i(t)\right\rangle=\sum_{i=1}^M c_i(t)\Op U_i(t) \left|\psi_i(0)\right\rangle, \label{ansatz}
\end{eqnarray}
where $M \le n$ and $\Op U_i(t)$ is a non unique time-dependent unitary transformation of the reference state $\psi_i(0)$. The simulation involves calculation of matrix elements of the Hamiltonian in the time-dependent basis.  An efficient computation of the matrix elements of the Hamiltonian  $\left\langle \psi(t)_i\left|\Op H\right|\psi(t)_j\right\rangle$ is possible only if  both the transformation 
\begin{eqnarray}
\Op H\rightarrow \Op U_i(t)^{\dagger}\Op H \Op U_j(t). \label{cond}
\end{eqnarray}
and the computation of matrix elements
\begin{eqnarray}
\left\langle \psi_i(0)\left|\Op H\right|\psi_j(0)\right\rangle \label{matrixel}
\end{eqnarray}
can be performed efficiently.

Let operators $\left\{\Op X_i\right\}$ span the SGA of the system. Then the Hamiltonian can be represented as  a polynomial in $\Op X_i$,  as in Eq.(\ref{bhlie}). 
If the unitary transformation $\Op U_j(t)$ is generated by an element of the SGA, the transformation (\ref{cond}) can be performed group-theoretically \cite{Gilmore}, i.e., efficiently. It follows that if each $\psi_i(t)$ in Eq.(\ref{ansatz}) can be represented as an orbit of the reference state $\psi_i(0)$ under the action of the SGA:
\begin{eqnarray}
\psi_i(t)&=&e^{i \sum_i \alpha_i(t) \Op X_i}\psi_i(0), \label{orbit}\\
\Op X_i&\in&SGA,
\end{eqnarray}
the transformation (\ref{cond}) can be performed efficiently.
Eq.(\ref{orbit}) defines a \textit{generalized coherent state} (GCS), associated with the SGA and the reference state $\psi_i(0)$ \cite{Perelomov,Gilmore}. The choice of the time-dependent computational basis of GCS, associated with the SGA of the system, guarantees that the transformation (\ref{cond}) can be performed efficiently. 

Complexity of computing matrix elements (\ref{matrixel})  generally depends on the Hilbert space representation of the SGA. But if the reference state $\psi_i(0)$ is a maximal (minimal) weight state of the representation, the computation can be performed group-theoretically.
These considerations suggest the choice of the basis of the GCS, associated with the SGA of the system and a maximal (minimal) weight states of the representation, as the time-dependent computational basis of the computation.
\subsubsection{The choice of the target observables for the simulation}
\label{sec:TheChoiceOfTheSGAObservablesForTheSimulation}
The choice of computational  basis of  GCS, associated with the SGA of the system and a maximal (minimal) weight state of the representation, allows  computing  matrix elements $\left\langle \psi(t)_i\left|\Op X\right|\psi(t)_j\right\rangle$ of an observable $\Op X$ efficiently if $\Op X$ is a polynomial in $\Op X_i$. Such observables  can be used as target observables for the efficient simulation.
Elements $\Op X_i$ of the SGA are the simplest choice of the target observables.  
\subsubsection{Solving  the stochastic Nonlinear Schr\"odinger Equation}
\label{sec:SolvingTheStochasticNonlinearSchroedingerEquation}
The size $M$ of the computational basis  in an efficient simulation must be independent of the Hilbert space dimension.  Generically a unitarily evolving system will evolve into a superposition of  $M\sim n$ GCS. Therefore, generically, a unitary dynamics cannot be simulated efficiently.
The open-system dynamics, corresponding to the weak measurement of the SGA operators of the evolving system, is simulated by averaging over stochastic pure-state realizations  of the measurement. Each realization is governed by the stochastic Nonlinear Schr\"odinger equation (sNLSE)\cite{gisin84,diosi88,gisin92}. Solution of the sNLSE, can be expressed in the form Eq.(\ref{ansatz}) with $M \ll n$, provided the  measurement is sufficiently strong. 
\subsubsection{Averaging}
\label{sec:Averaging}
Expectation values of the SGA observables are calculated in each stochastic realization of the sNLSE. Averaging over the realizations recovers the open-system dynamics of the observables. The number of realizations needed to compute the averages to a given relative error is independent of the Hilbert space dimension. Therefore, the averaging can be performed efficiently.
\subsubsection{Convergence}
\label{sec:ConvergenceToTheUnitaryEvolution}
Numerical solution of the sNLSE converges as the size $M$ of the computational basis  is increased.
The open-system dynamics of the SGA observables converges to the original unitary dynamics as the strength $\gamma$ of the measurement is decreased. The smaller $\gamma$ the larger $M$ is needed for the convergent simulation.
If a certain generic condition on the SGA and its Hilbert space representation is satisfied \cite{khasin081},  the open system follows the evolution on widely separated time-scales.  Due to this time-scales separation  the open-system dynamics of the SGA observables converges to the unitary evolution at the value of  $\gamma$ which is sufficiently large to ensure that $M \ll n$ in the expansion (\ref{ansatz}) of the convergent pure-state solution of the corresponding sNLSE. 

\subsection{A linear implementation}
\label{sec:ALinearImplementation}
 The algorithm outlined in the previous subsection can be implemented in a variety of ways. This is  in part due to the non-uniqueness of the transformations $\Op U_i(t)$  of the computational basis elements in Eq.(\ref{ansatz}).  In the present implementation we adopt the choice $\Op U_i(t)=\Op  U_j(t)=\Op U(t)$, termed the \textit{linear implementation}. In the linear implementation the pure-state solution of the sNLSE is represented by
 \begin{eqnarray}
\left|\Psi(t)\right\rangle=\sum_{i=1}^M c_i(t)\Op U(t) \left|\psi_i(0)\right\rangle, \label{ansatz1}
\end{eqnarray}
 where $\psi_i(0)$ is a fixed set of the GCS, associated with the SGA and a maximal (minimal) weight state of the Hilbert space representation of the algebra.
 
The advantage of this choice is that  updating of the computational basis is performed by a linear (unitary) transformation, leading to high numerical stability.  The disadvantage is a limited flexibility of the computational basis evolution, which may result in a higher computational cost.

 An infinitesimal time step of the simulation can be represented by a superposition of the nonlinear transformation $\Op T_1$ of the state represented in the fixed instantaneous basis, i.e., the transformation of the vector $c_i$ in Eq.(\ref{ansatz1}), and a subsequent unitary transformation $\Op T_2$ of the basis itself. The transformation $\Op T_1$  corresponds to the  evolution of the state, driven by the sNLSE. The  transformation 
$\Op T_2$   corresponds to the updating of the computational basis and ensures that the evolving state stays in the Hilbert subspace spanned by the instantaneous basis. Connection of the infinitesimal unitary transformation $\Op T_2$ to $\Op U(t)$ in Eq.(\ref{ansatz1}) is established by the following relation:
\begin{eqnarray}
T_2=\Op 1-\Op U(t)'dt. \label{tr2}
\end{eqnarray}

The transformations $T_1$ and $T_2$ can be  given a simple geometric interpretation in the special case of the ${\mathfrak{su}}(2)$ SGA spanned by the operators $\Op J_x$, $\Op J_y$ and $\Op J_z$, satisfying the commutation relations 
\begin{eqnarray}
\left[\Op J_i,\Op J_j\right]=i \epsilon_{ijk} \Op J_k .\label{commutation}
\end{eqnarray}
The GCS of the ${\mathfrak{su}}(2)$, termed spin-coherent states, can be represented by points in the phase space, associated with the algebra \cite{arecchi,viera}. The phase-space  is a two-dimensional sphere and the GCS can be parametrized by spherical coordinates. The  $2j+1$ dimensional Hilbert space, carrying an irreducible representation of the algebra, is spanned by a linearly independent  basis of GCS, represented by $2j+1$  points on the sphere. A  localized state can be expanded in a small fraction of the total basis, represented by points in the interior of a relatively small area of the phase-space. 

A single realization of the weak measurement of operators $\Op J_x$, $\Op J_y$ and $\Op J_z$ of a quantum system, evolving under the action of a Hamiltonian $\Op H$ is described by the following sNLSE:
\begin{eqnarray}
d\left|{\psi}\right\rangle &=&-i \Op H \ dt \left|{\psi}\right\rangle \nonumber \\ 
&-&\gamma \sum_{i=1}^3  \left( \Op J_i-\left\langle \Op J_i\right\rangle_{\psi}  \right)^2 dt \left|{\psi}\right\rangle+ \sum_{i=1}^3\left(  \Op J_i-\left\langle \Op J_i\right\rangle_{\psi}\right) d\xi_i \left|{\psi}\right\rangle, \label{snlse}
\end{eqnarray} 
where $\left\langle \Op J_i\right\rangle_{\psi}$ are the instantaneous expectation values of the single-particle operators and the Wiener fluctuation terms $d\xi_i$ satisfy 
\begin{eqnarray}
\left[d\xi_i\right]=0, \ \ \ \left[d\xi_id\xi_j\right]=2\gamma \delta_{ij}dt, \label{noise}
\end{eqnarray}
where $\left[ \ \right]$ denotes statistical averaging. 
Parameter $\gamma$ is termed the \textit{strength of  measurement} in what follows.
 
Hamiltonian $\Op H$ nonlinear in  $\Op J_x$, $\Op J_y$ and $\Op J_z$  generates  delocalization of an initially localized state in the phase-space. The non-unitary stochastic contribution of the evolution due to the measurement, represented by the last two terms in Eq.(\ref{snlse}), leads to  localization of the state \cite{khasin082}. A single   realization of the open system evolution can be visualized in the phase-space as a stochastic trajectory of  finite width.

The infinitesimal nonlinear transformation $\Op T_1$ 
\begin{eqnarray}
\Op T_1=\Op I+\left\{-i \Op H dt -\gamma \sum_{i=1}^3  \left( \Op J_i-\left\langle \Op J_i\right\rangle_{\psi}  \right)^2 dt+ \sum_{i=1}^3\left(  \Op J_i-\left\langle \Op J_i\right\rangle_{\psi}\right) d\xi_i\right\}, \label{t1exp}
\end{eqnarray}
includes two parts: the original Hamiltonian part and the non-unitary part, which contains  a stochastic and a deterministic element. Both elements of  the non-unitary part depend on the instantaneous state of the system. 

The transformation $T_1$ generates a drift of the state in the phase-space. The computational basis must be updated to compensate for the drift.
The GCS basis is updated at each time step by performing the unitary transformation $T_2$, Eq.(\ref{tr2}). Practically, a different but physically equivalent transformation has been found to be more convenient. Instead of updating the basis to follow the evolving state we update the state to match the fixed  basis. Therefore, the inverse unitary transformation 
$\Op T_2^{\dagger}$ is performed on the state itself and the  transformation $\Op T_2$ is applied to the Hamiltonian and other observables. The unitary transformation $\Op T_2^{\dagger}$ is generated by the ${\mathfrak{su}}(2)$  elements $\Op J_x$, $\Op J_y$ and 
$\Op J_z$ and rotates the state to the  position localized
in the fixed basis of the GCS. 

For example, let the initial state be a spin-coherent state, represented by  the south pole of the phase space. The expectation values of the  operators $\Op J_x$ and $\Op J_z$ vanish in this state. An infinitesimal transformation $\Op T_1$ takes the state into a new position having finite expectation values of  $\Op J_x$ and $\Op J_z$. The purpose of the unitary transformation $\Op T_2^{\dagger}$ is to rotate the state back to the state where the expectation values of  $\Op J_x$ and $\Op J_z$ vanish. This rotation is performed by the $SU(2)$ group transformation and is equivalent to rotating  the density operator in the Hilbert-Schmidt space to the state where the projection of the  density operator on the SGA is spanned by the $\Op J_z$ operator. This procedure is analogous to the Schmidt-decomposition of the state in the generalized entanglement setting \cite{viola03}.

The back-rotation of the operators by the transformation  $\Op T_2$ can be performed analytically, knowing how the ${\mathfrak{su}}(2)$ algebra elements transform under the $SU(2)$ group itself.
The superposition of the two transformations $\Op T_2^{\dagger} \circ \Op T_1$ of the state results in a new state,  localized about the south pole. 
\section{Application to the two-site Bose-Hubbard model}
\label{sec:ApplicationToTheTwoSiteBoseHubbardModel}
\subsection{Introduction}
\label{sec:Introductionbh}
The gas of cold bosonic atoms in a double-well trap is described by the two-site Bose-Hubbard model, Eq.(\ref{bhgenral}). The Hamiltonian of the system
 contains two terms: the one-body  term responsible for hopping of the atoms from one well to another and the two-body or the interaction term responsible for the on-site attraction or repulsion of the atoms. Here we  consider repulsive interaction, measured by the coupling strength $2U/N$, $U>0$, where $N$ is the total number of particles in the trap. The hopping rate is determined by the coefficient $\Delta$ of the one-body part.

The SGA of the system is the ${\mathfrak{su}}(2)$ algebra of the single-particle operators (\ref{suops}). The system of $N$ particles correspond to the $N+1$-dimensional irreducible representation of the algebra with the principle pseudo-spin quantum number $j=N/2$.
Condensed state of the system is a spin coherent state \cite{vardi00,trimborn}. 

\begin{figure}[t]
\epsfig{file=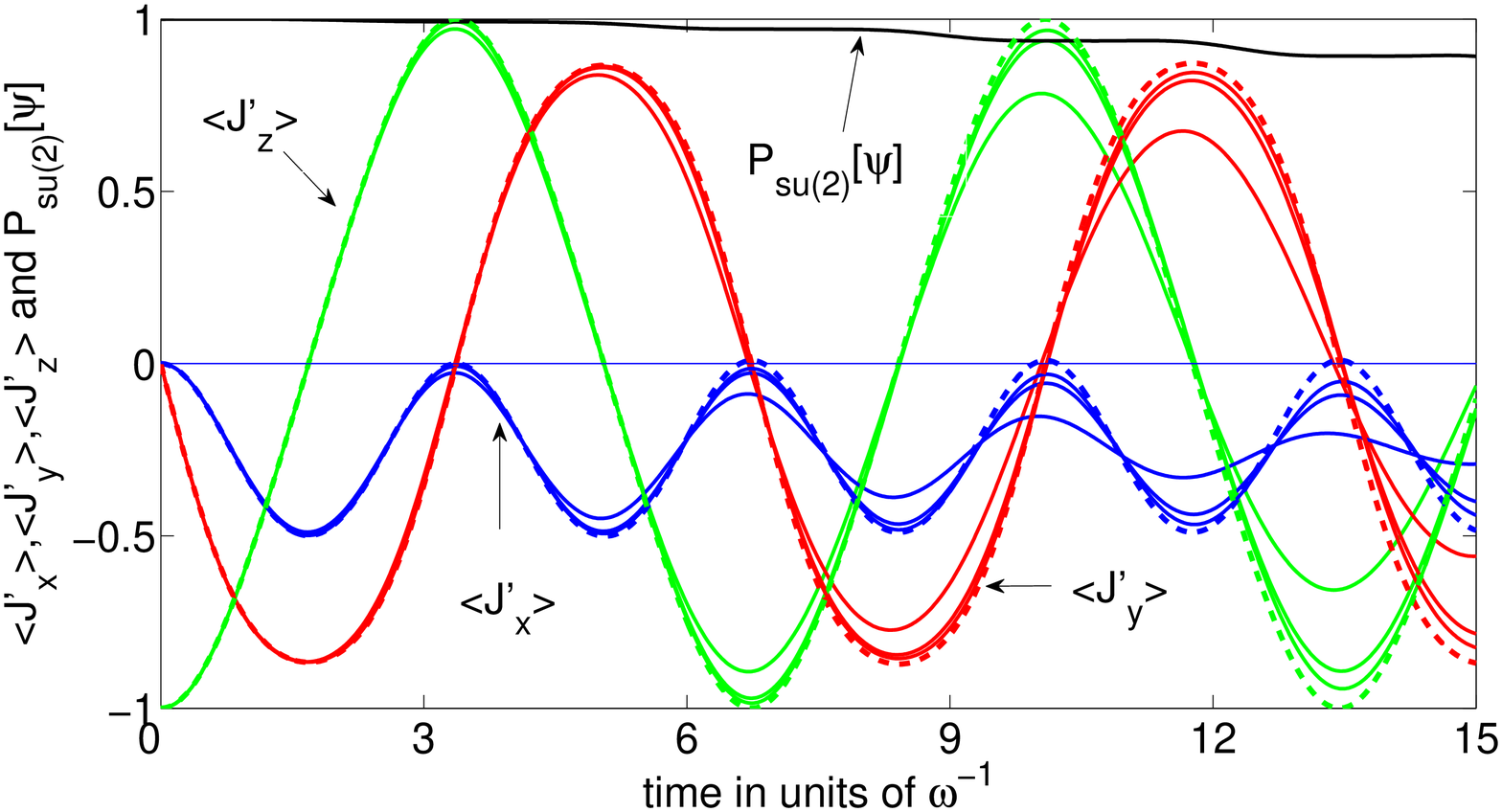, width=16.0cm, clip=} 
\caption{ (Color online) Mean-field (dashed lines) vs. exact (solid lines) solutions of the Schr\"odinger equation, corresponding to the Hamiltonian (\ref{bhlie}). Expectation values of the target observables $\Op J'_x=\frac{2}{N}\Op J_x$ (blue), $\Op J'_y=\frac{2}{N}\Op J_y$ (red) and $\Op J'_z=\frac{2}{N}\Op J_z$ (green), Eq.(\ref{suops}), are plotted  as  functions of time for $N=128$, $N=256$ and $N=512$ atoms. The generalized purity (black), Eq.(\ref{gpurity}), is plotted  in $N=512$ case.
The system is prepared in a spin-coherent state corresponding to the condensate placed in the left well of the trap. The value of the on-site interaction is chosen $U/\omega =1$. Time is measured in units of $\omega^{-1}$. As the number of particles grows the exact solution approaches the mean-field solution.}
\label{fig:meanweak}
\end{figure}

The character of dynamics  depends on a single parameter $U/\omega$, where $\omega=2\Delta$. At $N \gg 1$ the  weak interaction regime $U/\omega \le 1$, corresponds to dynamics preserving coherence of the atomic condensate for times of order $\sqrt{N}$ \cite{anglin00}. High coherence corresponds to a weak delocalization of an initial spin-coherent state.  The corresponding dynamics is very accurately described  by  the mean-field solution of the Schr\"odinger equation [Cf. Fig. \ref{fig:meanweak}], assuming a spin-coherent form of the evolving state \cite{kramer}.
\begin{figure}[t]
\epsfig{file=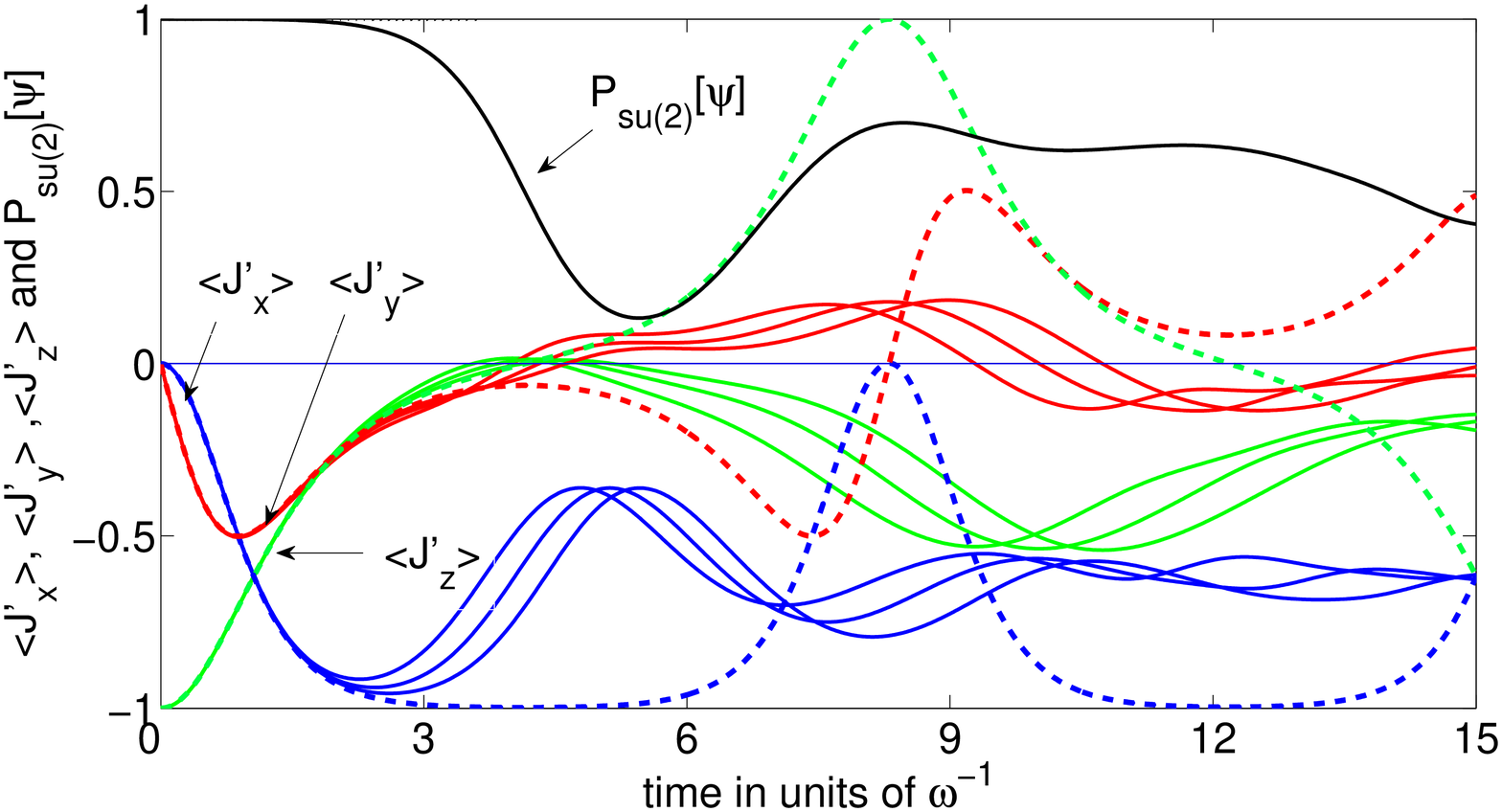, width=16.0cm, clip=} 
\caption{ (Color online) The same as in Fig.\ref{fig:meanweak} but for $U/\omega=2$. The mean-field approximation breaks down.}
\label{fig:meanstrong}
\end{figure}

When $U/\omega > 1$ the phase-space picture of the mean-field dynamics changes: an unstable fixed point appears and the associated separatrix  \cite{vardi00}. The mean-field solution breaks down in the vicinity of the separatrix. As a consequence, an initial spin-coherent state prepared in the vicinity of the separatrix is expected to evolve into a delocalized state. The corresponding dynamics is  characterized by the exponential loss of coherence \cite{khodor} as measured by the purity of the single-particle density operator. The delocalization corresponds to depletion of the condensate. This parametric regime poses the real challenge for the simulation and is chosen to test the algorithm. 

The loss of coherence can  be measured by the generalized purity \cite{viola03} of the state with respect to the ${\mathfrak{su}}(2)$ algebra of the single-particle observables $\Op J_x$, $\Op J_y$ and $\Op J_z$.  The generalized purity of the state $\psi$ is defined by
\begin{eqnarray}
P_{{\mathfrak{su}}(2)}[\psi]\equiv \frac{4}{N^2}\sum_{i=x,y,z}\left\langle \Op J_i \right\rangle^2. \label{gpurity}
\end{eqnarray}
The generalized purity equals  unity if  $\psi$ is a spin-coherent state and is less than unity otherwise. Since the mean-field state is  spin-coherent  by construction, the mean-field approximation breaks down if the generalized  purity decreases. 

Figs. \ref{fig:meanweak} and \ref{fig:meanstrong} show  dynamics of the system of $N=128$, $N=256$ and $N=512$ atoms corresponding to $U/\omega = 1$ and $U/\omega = 2$ respectively, obtained by solving the Schr\"odinger equation using the Chebyshev method \cite{k28}. For initial state of the system we choose a condensate of  atoms prepared in the left well. It is a spin-coherent eigenstate of the operator $\Op J_z$. The expectation value of $\Op J_z$ measures the difference of the right- and left-well populations of the atoms. The mean-field solution works well for the weak interaction regime $U/\omega = 1$, where the generalized purity of the evolving remains close to unity. In the case of  $U/\omega = 2$ the same initial  state follows  dynamics in the vicinity of the separatrix in the mean-field picture \cite{vardi00}. As a consequence, the generalized purity decreases on the time-scale $\omega^{-1} \ln N$ following the depletion of the condensate and the mean-field solution fails to reproduce correctly the character of the dynamics.

\subsection{Application to a system of large number of atoms}
\label{sec:ApplicationToALargeNumberOfParticles}
\subsubsection{Equations of motion}
\label{sec:EquationsOfMotion}
The numerical solution of the sNLSE (\ref{snlse}) proceeds as a sequence of infinitesimal transformations:
\begin{eqnarray}
\left|\psi\right\rangle &\rightarrow& \Op T_1\left|\psi\right\rangle \label{t1}\\
\left|\psi\right\rangle &\rightarrow& \Op T_2^{\dagger}\left|\psi\right\rangle, \label{t2} \\
\Op X &\rightarrow& \Op T_2^{\dagger} \Op X \Op T_2, \label{t3}
\end{eqnarray}
where $\Op X$ stands for all relevant operators, i.e., the Hamiltonian (\ref{bhlie}) and the target observables (\ref{suops}). The state $\psi$ in Eqs.(\ref{t1}) and (\ref{t2}) is given by
 \begin{eqnarray}
\left|\Psi(t)\right\rangle=\sum_{i=1}^M c_i(t)\left|\psi_i(0)\right\rangle, \label{ansatz2}
\end{eqnarray}
where $\psi_i(0)$ is the \textit{fixed} computational basis of spin-coherent states.
From Eqs.(\ref{t1}), (\ref{t1exp}) and (\ref{ansatz}) we obtain
\begin{eqnarray}
\sum_{k=1}^M &C_{jk}&dc_k \label{motion}\\
&=&\sum_{k=1}^M \left\{-i (\Op H)_{jk} dt -\gamma \sum_{i=1}^3  \left( (\Op J_i)_{jk}-\left\langle \Op J_i\right\rangle_{\psi}  \right)^2 dt+ \sum_{i=1}^3\left(  (\Op J_i)_{jk}-\left\langle \Op J_i\right\rangle_{\psi}\right) d\xi_i\right\}c_k, \nonumber 
\end{eqnarray}
where $(\Op X)_{jk}\equiv \left\langle \psi_j(0)\left|\Op X\right|\psi_k(0)\right\rangle$ and 
\begin{eqnarray}
C_{jk}\equiv  \left\langle \psi_j(0)|\psi_k(0)\right\rangle
\end{eqnarray}
is the overlap matrix. It should be noted that the overlap matrix is time-independent.

The expectation values of the SGA observables are calculated in each realization (\ref{ansatz2}) of the weak measurement:
\begin{eqnarray}
\left\langle \Op J_i \right\rangle_{\psi}\equiv\left\langle \Psi(t)\right|\Op J_i\left|\Psi(t)\right\rangle=\sum_{k,j}^M c_j^*(t)c_k(t)\left( \Op J_i\right)_{jk}.
\end{eqnarray}
Averaging over the realizations gives the expectation values in the surrogate open-system dynamics (\ref{lvn})
\begin{eqnarray}
\left\langle \Op J_i \right\rangle=\left[ \left\langle \Op J_i \right\rangle_{\psi}\right]. \label{finalexp}
\end{eqnarray}
Convergence of the simulation is checked on the level of Eq.(\ref{finalexp}).
\subsubsection{Choice of numerical parameters}
\label{sec:TheChoiceOfNumericalValues}

A converged solution, Eq.(\ref{finalexp}), is independent of the size $M$ of the computational basis, of the strength of the measurement $\gamma$ and of the size of the time-step $dt$ in the Eq.(\ref{motion}). The convergence is obtained by increasing $M$  and decreasing  $dt$ and $\gamma$.

\paragraph{Choice of the basis.}
\label{sec:ChoiceOfBasis}
A spin coherent state can be represented by a point on the phase-space sphere. It is expected that a state having high generalized purity  can be represented by a superposition of spin-coherent states localized in a small area of the phase-space.  
Solution of the sNLSE for an initial spin-coherent state has high generalized purity if the measurement  is sufficiently strong \cite{khasin081,khasin082}. Therefore, a localized computational basis is expected to span the solution. Representation in terms of spin-coherent states is non-unique in general and the choice of the localized basis is non-unique, too. As a guideline for the choice of the basis we have used the following consideration. 
 A set of $N+1$ spin-coherent states comprises the full basis of the Hilbert space, which can be represented by a  distribution of $N+1$ points on the sphere.  A natural choice is a uniform distribution. Then a localized computational basis of $M$ states can be mapped into $M$ points uniformly distributed over an area $A$ of the phase-space unit sphere, with $A=4\pi M/(N+1)$. This mapping determines the parametrization of the computational basis.
 
The value of $M$ in the convergent solution of the sNLSE depends on the strength of  measurement $\gamma$. It can be estimated balancing the  rate of the delocalization, induced by the Hamiltonian, and the rate of the localization, induced by the measurement. The delocalization rate in the state prepared in the vicinity of the mean-field separatrix with $U/\omega=2$ is of the order of $\omega$ \cite{khodor}. The  localization rate in the state having generalized purity $P_{{\mathfrak{su}}(2)}[\psi]$, is of the order of $\gamma \sum_{i=1}^3 \left( \left\langle  \Op J_i^2\right\rangle-\left\langle \Op J_i\right\rangle^2\right)\sim \gamma N^2 (1-P_{{\mathfrak{su}}(2)}[\psi])$ \cite{khasin081}, which can be brought to the form  $\gamma M N$ using  relation $M \lesssim  N \left(1-\sqrt{P_{{\mathfrak{su}}(2)}[\psi]}\right)$ derived in Ref.\cite{khasin081}. Equating the rates we obtain
\begin{eqnarray}
M \lesssim \frac{\omega}{\gamma N}. \label{mestimate}
\end{eqnarray}

As $\gamma$ goes to zero $M$ approaches its value in the corresponding unitary dynamics, which in  the strong interaction regime may become of the order of $N$.
The convergent solution of Eq.(\ref{lvn}) is obtained at $\gamma \ll \omega/(N \ln N)$ [Sec. (\ref{sec:ChoiceOfGamma})]. The value of $M$ in (\ref{mestimate}) is determined by $\gamma_{\epsilon}\equiv\epsilon \omega/(N \ln N)$, $\epsilon \ll 1$. Therefore,
\begin{eqnarray}
M \lesssim \frac{\omega}{\gamma_{\epsilon} N}\sim \epsilon^{-1}\ln N.\label{mestimate1}
\end{eqnarray}
\paragraph{Choice of the time-step.}
\label{sec:ChoiceOfTimestep}
 Eq.(\ref{t1exp}) implies that the size of the time-step should be chosen to satisfy  $||\Op H  \left|\psi\right\rangle ||\ll dt^{-1}$. Since adding a (time-dependent) phase to the state does not change the expectation values, the Hamiltonian can be substituted for $\Op H -\left\langle \Op H\right\rangle_{\psi}$ in the equations of motion.   The condition of the size of the time-step becomes $\left\|\left(\Op H - \left\langle \Op H\right\rangle_{\psi} \right)\left|\psi\right\rangle \right\|=\left(\left\langle \Op H^2 \right\rangle_{\psi}-\left\langle \Op H \right\rangle_{\psi}^2\right)^{1/2}\ll dt^{-1}$, i.e., the size of the time-step must be smaller than the uncertainty of the Hamiltonian. The uncertainty of the Hamiltonian can be estimated in a localized state $\psi$.  The phase-space representation of the Hamiltonian  quadratic in the operators $\Op J_i$, as in Eq.(\ref{bhlie}),  varies on the scale of unity, whereas the width of the phase-space representation of the localized state for $M \ll N$ is $\sim \sqrt{M/N} \ll 1$. As a consequence, the uncertainty of the Hamiltonian is of the order of $\left\langle \Op H\right\rangle_{\psi} \sqrt{M/N} \lesssim \omega \sqrt{MN}$
and the time-step must satisfy
\begin{eqnarray}
\omega dt \ll \left(N M\right)^{-1/2}. \label{timestep}
\end{eqnarray}
A similar estimation of other terms in Eq.(\ref{t1exp}), using  condition $\gamma \ll \omega/(N \ln N)$  for a convergent simulation [Sec. (\ref{sec:ChoiceOfGamma})] and Eqs.(\ref{noise}) , leads to 
\begin{eqnarray}
\omega dt \ll \frac{ \ln N}{M}, \label{secondterm}
\end{eqnarray}
Since inequality (\ref{timestep}) is stronger than inequality (\ref{secondterm}),  the scaling of the size of time-step of the simulation is determined by inequality (\ref{timestep}). As a consequence, the time of computation scales as $\sqrt{n}$, where $n=N+1$ is the Hilbert space dimension of the system (Cf. Fig. \ref{fig:time}).
\paragraph{Choice of $\gamma$.}
\label{sec:ChoiceOfGamma}
In order that the effect of measurement on the dynamics of the target observables $\Op J_x$, $\Op J_y$ and $\Op J_z$ can be neglected, it is  sufficient that the contribution of the measurement to the growth of the total uncertainty $\sum_{i=1}^3 \left( \left\langle  \Op J_i^2\right\rangle-\left\langle \Op J_i\right\rangle^2\right)$ in the open-system evolution (\ref{lvn}) is negligible  on the 
time-scale of the simulation, compared to the  uncertainty  of the initial spin-coherent state. 

To estimate the contribution of the measurement to the growth of total uncertainty we consider dynamics of the single-particle observables under the action of measurement alone  \cite{khasin081}:
\begin{eqnarray}
\left\langle \Op J_i(t)\right\rangle=\left\langle \Op J_i(0)\right\rangle e^{-2 \gamma t}. \label{meas}
\end{eqnarray}
The corresponding evolution of the total  uncertainty is
\begin{eqnarray}
 \Delta(t)&=&\sum_{i=1}^3 \left( \left\langle  \Op J_i^2\right\rangle-\left\langle \Op J_i\right\rangle^2\right)=\frac{N}{2}\left(\frac{N}{2}+1\right)-\sum_{i=1}^3  \left\langle J_i(0)\right\rangle^2 e^{-4\gamma t}\nonumber \\ &=&\frac{N}{2}\left(\frac{N}{2}+1\right)-\frac{N}{2}^2 e^{-4\gamma t}. \label{delta}
\end{eqnarray}
The uncertainty in a spin-coherent state is minimal:  $\Delta_{min}=N/2$. The effect of measurement can be neglected on a time-interval $\tau$ if 
 \begin{eqnarray}
 \Delta(\tau)-\Delta_{min} \ll \Delta_{min},\label{deltacon}
\end{eqnarray}
From Eqs.(\ref{delta}) and (\ref{deltacon}) 
  \begin{eqnarray}
 \frac{N}{2}\left(\frac{N}{2}+1\right)-\frac{N}{2}^2 e^{-4\gamma \tau }-\frac{N}{2}&=&\frac{N}{4}^2\left(1-e^{-4\gamma \tau}\right)\ll \frac{N}{2} \ \ \Rightarrow\nonumber \\
 \gamma \tau N^2&\ll& \frac{N}{2}\ \ \Rightarrow \nonumber \\
 \gamma &\ll&  \left({2 N\tau}\right)^{-1},\label{radius}
\end{eqnarray}
where assumption $1 \ll N$ is used. 
For the simulation on the time interval of order $\omega^{-1} \ln N$  condition (\ref{radius}) becomes
\begin{eqnarray}
\gamma \ll \frac{\omega}{N \ln N}. \label{radius1}
\end{eqnarray}
Therefore, the open-system dynamics of the target observables converges to the original unitary evolution at the strength of measurement $\gamma_{\epsilon}=\epsilon \frac{\omega}{N \ln N}$, $\epsilon \ll 1$.
\subsubsection{Numerical results}
\label{sec:NumericalResults}
\begin{figure}[t]
\epsfig{file=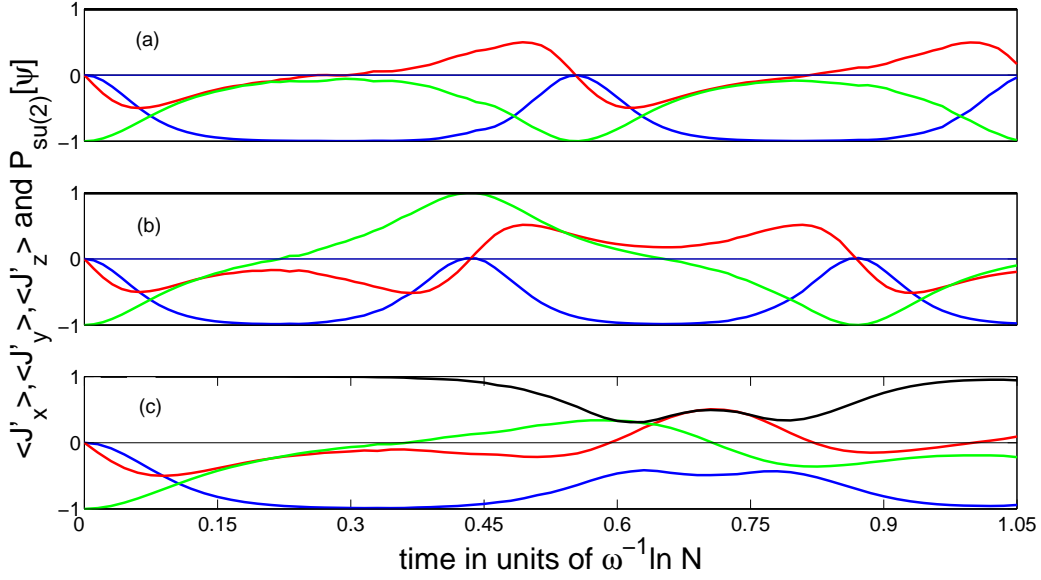, width=16.0cm, clip=} 
\caption{ (Color online). Panels (a) and (b) display solutions of  sNLSE (\ref{snlse}), corresponding to two different realizations of the weak measurement. The normalized expectation values of the target observables $\Op J'_x=\frac{2}{N}\Op J_x$ (blue), $\Op J'_y=\frac{2}{N}\Op J_y$ (red) and $\Op J'_z=\frac{2}{N}\Op J_z$ (green), Eq.(\ref{suops}) and the generalized purity (black), Eq.(\ref{gpurity}), are plotted as  functions of time.  The system of $N=2\cdot 10^4$ atoms is prepared in a spin-coherent state corresponding to the condensate placed in the left well of the trap.   The value of the on-site interaction is chosen $U/\omega=2$. Time is measured in units of $\omega^{-1}\ln N$. The size of the time-dependent basis  $M=60$.
Panel (c) displays the average of the two realizations and the corresponding generalized purity.}
\label{fig:singles}
\end{figure}
\begin{figure}[t]
\epsfig{file=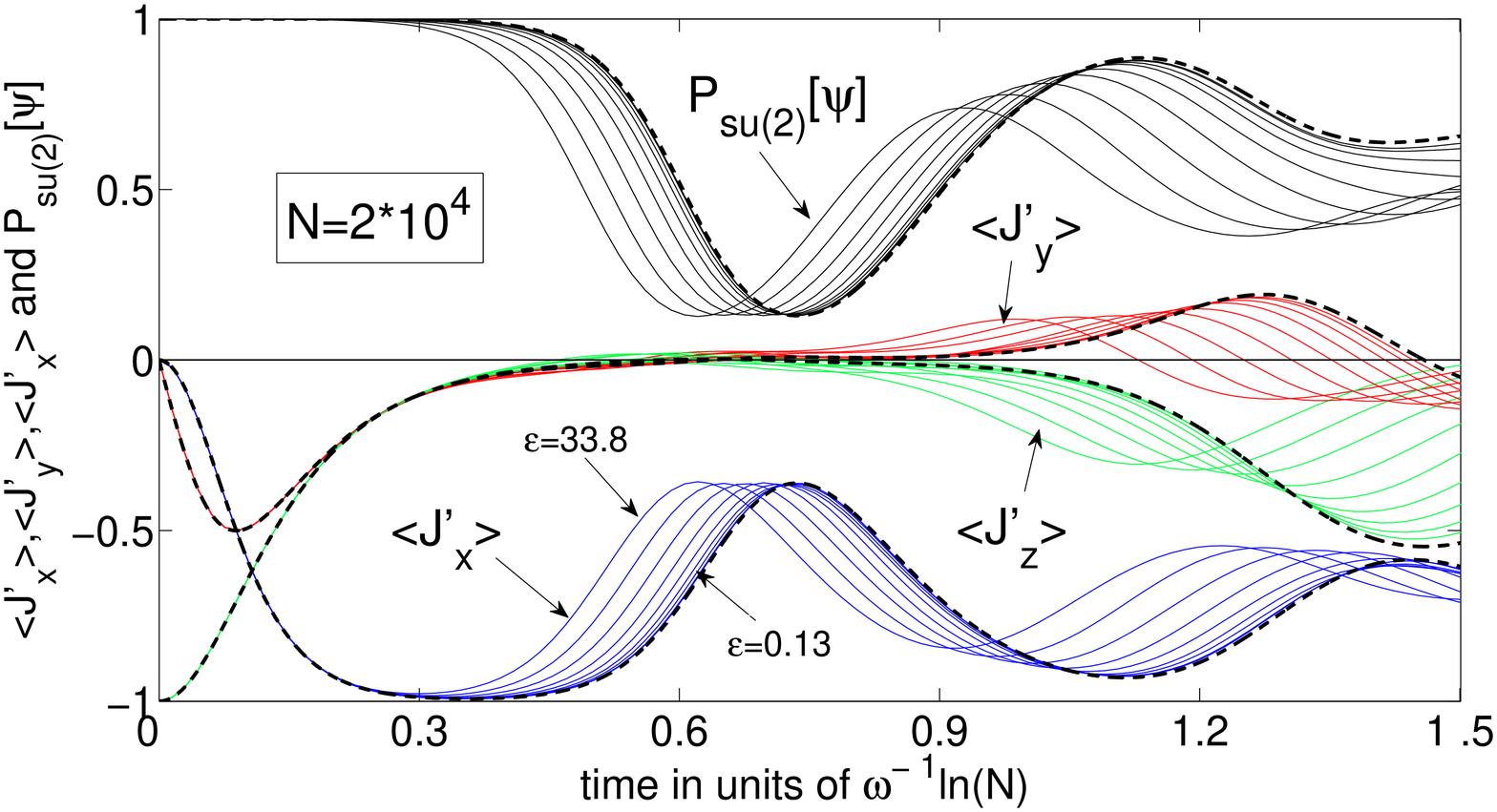, width=16.0cm, clip=} 
\caption{ (Color online). $N=2\cdot 10^4$. Expectation values of the  the target observables $\Op J'_x=\frac{2}{N}\Op J_x$ (blue solid), $\Op J'_y=\frac{2}{N}\Op J_y$ (red solid) and $\Op J'_z=\frac{2}{N}\Op J_z$ (green solid), Eq.(\ref{suops}), averaged over $10^4$   realizations of  the sNLSE (\ref{snlse}) for  the set of  values $\{0.13, 0.26, 0.53, 1.06, 2.11, 4.23, 8.45,16.9, 33.8\}$ of $\epsilon\equiv\gamma N \ln N/\omega$.  The expectation values and the corresponding generalized purity (black), Eq.(\ref{gpurity}), are plotted as functions of time.  The initial state of the system and the value of $U/\omega$ are the same  as in Fig.(3).  Time is measured in units of $\omega^{-1}\ln(N)$.  The results of the exponentially convergent computation using sparse-matrix Chebyshev techniques  are presented for comparison (black dashed).}
\label{fig:final}
\end{figure}

 As an application of the algorithm we have performed the simulation of the system of $N=2 \cdot 10^4$ and $N=2 \cdot 10^6$ atoms. Panels (a) and (b) in Fig. (\ref{fig:singles}) display dynamics of the target observables for two different realizations of the weak measurement governed by sNLSE (\ref{snlse}). Each solution is converged with respect to the size $M$ of the time-dependent basis.
The size of the basis is $M=60$ and the value of $\gamma$ is $0.3\omega/(N \ln N)$. The generalized purity (\ref{gpurity}) stays very close to unity as a consequence of the localization. The evolution of the expectation values of  target observables in the two realizations of the weak measurement are identical   on the  time interval $0\le t\le 0.15 \omega^{-1} \ln N$, corresponding to high generalized purity of the exact solution [Cf. Fig.\ref{fig:final}]. On this time interval the mean-field solution [Cf. Fig.\ref{fig:meanstrong}] is a  perfect approximation. At $t \gtrsim 0.3 \omega^{-1} \ln N$ the two realizations start to diverge. The computation of each realization takes about $30$ seconds on a PC.

 Panel (c) shows the  dynamics of the target observables averaged over the two realizations and the corresponding generalized purity. An important observation is the decrease of the generalized purity.  It implies the loss of the generalized purity in the exact dynamics  and gives an estimate of the delocalization time-scale. We conclude that some features of a many-body dynamics can be inferred from a small number of realizations.
   \begin{figure}[t]
\epsfig{file=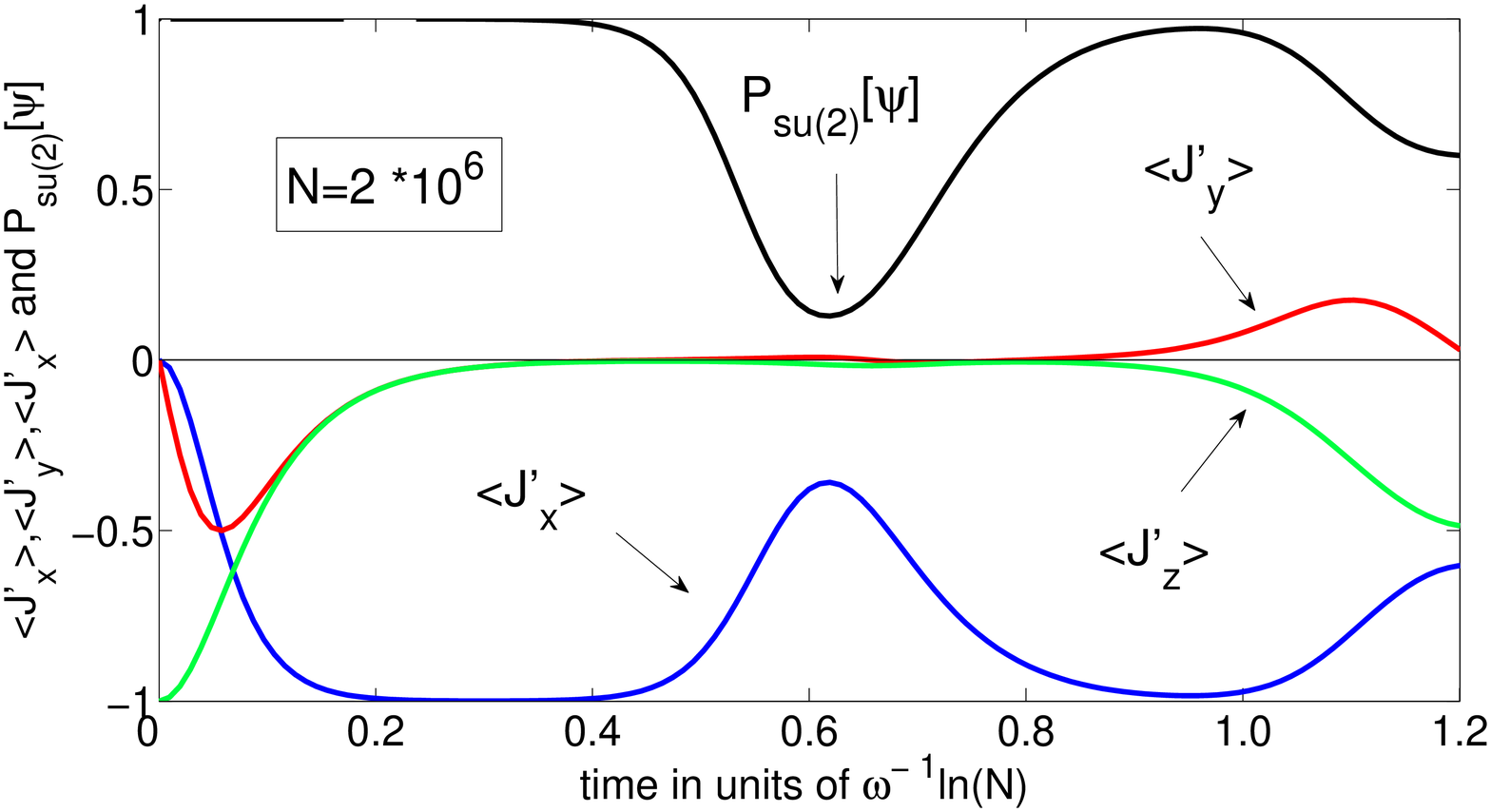, width=16.0cm, clip=} 
\caption{ (Color online). $N=2\cdot 10^6$. Expectation values of the  the target observables $\Op J'_x=\frac{2}{N}\Op J_x$ (blue), $\Op J'_y=\frac{2}{N}\Op J_y$ (red) and $\Op J'_z=\frac{2}{N}\Op J_z$ (green) and the corresponding generalized purity (black) as functions of time. The initial state of the system and the value of $U/\omega$ are the same  as in Fig.(3). The simulation is converged to  $\sim 1.5 \%$, corresponding to averaging over $5000$ realizations of the sNLSE (\ref{snlse}). The size of the computational basis $M=90$.}
\label{fig:million}
\end{figure}
   \begin{figure}[t]
\epsfig{file=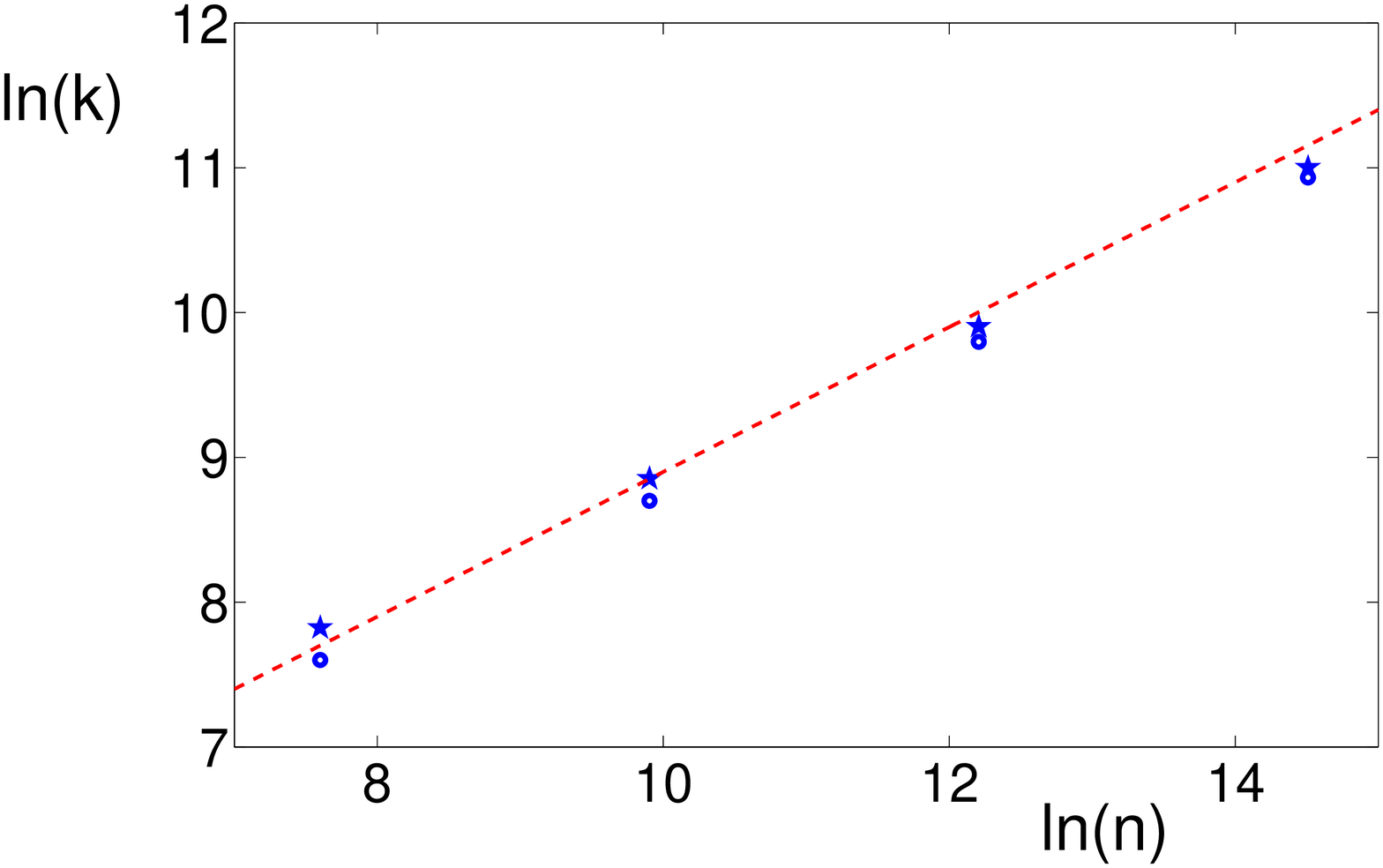, width=16.0cm, clip=} 
\caption{ (Color online). Empirical scaling of the minimal number of time steps $k$ per unit time necessary for convergent simulation, as a function of the Hilbert space dimension $n=N+1$. The estimate is performed by decreasing the number of steps until the computation becomes unstable. For convergent simulation it is sufficient to take $5$ to $10$ times $k$ number of steps. The results correspond to the size of the computational basis $M=60$. Stars and circles correspond to two different methods of performing time discretization. The dashed line (red) is a fit  to $\ln(k)=0.5 \ln(n)+c$.}
\label{fig:time}
\end{figure}

Averaging over $10^4$ stochastic trajectories leads to a relative error of $\sim 1\%$  in the expectation values of the observables. Figure (\ref{fig:final}) presents the averaged results. The averaging has been performed over solutions of the sNLSE, corresponding to $\epsilon\equiv\gamma N \ln N/\omega$ in the range $0.1 < \epsilon < 35 $. The averages are seen to converge as $\epsilon$ decreases.  The evolution on the time interval $t \lesssim 0.3\omega^{-1}\ln N$ corresponds to the mean-field behavior. At $t \gtrsim 0.5 \omega^{-1}\ln N$ the generalized purity decreases. Results of the simulation are compared to  the exponentially convergent numerical solution \cite{shimshon}, obtained by the sparse-matrix Chebyshev propagation method \cite{k56}. The correspondence of the converged results of the two algorithms  stays within $1-2 \%$. 
The size of the basis for the convergent solution  on the interval $0<t<\omega^{-1}\ln N$ is $M_{sNLSE}=60$ and on the interval $0<t<1.5\omega^{-1}\ln N$ is $M_{sNLSE}=180$.
The number of the time-steps for the propagation on the interval $0<t<1.5\omega^{-1}\ln N$ is $1.5 \cdot 10^5$.  

Fig.\ref{fig:million} presents results of simulating the system of $N=2 \cdot 10^6$ atoms. The simulation is converged by averaging over $5 \cdot 10^3$ solutions of the sNLSE to $\sim 1.5 \%$. The size of the computational basis is $M=60$. The number of timesteps is $8 \cdot 10^5$.

The scaling of the number of time-steps with the the Hilbert space-dimension $n=N+1$ was checked by finding the minimal number of time-steps $k$ per unit time for fixed $M$, where the computation is still stable. The log-log plot for $M=60$ in Fig. \ref{fig:time} shows the $\sqrt{n}$ scaling of $k$. Convergence of the simulations is obtained for the number of time-steps about $5-10$ times $k$   per unit time.
For $N=2 \cdot 10^4$ the simulation  using the Chebyshev  method is faster  due to the necessary averaging over the measurement realizations in the our method. For $N=2 \cdot 10^6$ our method is by orders of magnitude more efficient.

\section{Conclusions}
\label{sec:Conclusionsalgo}
The details and the first application of
an algorithm for simulation of a many-body dynamics, generated by a ${\mathfrak{su}}(2)$-Hamiltonian, are reported. The algorithm implements  idea, proposed in Ref.\cite{khasin081},  of using a surrogate open-system dynamics for  simulation of a unitary evolution of the quantum many-body systems.
The algorithm is applied to simulation of  dynamics of $N=2\cdot 10^4$ and $N=2\cdot 10^6$ cold atoms in the double-well trap. The dynamics reflects a competition between the hopping rate of the atoms
from well to well and the two-body repulsive interaction between the particles. The single-particle observables of the system are simulated  and the convergence of the simulation is checked.

The simulation is based on numerical solution of the stochastic Nonlinear Schr\"odinger Equation (sNLSE), which governs a single realization of the surrogate open-system dynamics. The open-system dynamics can be interpreted as a weak measurement of the elements of  spectrum-generating  ${\mathfrak{su}}(2)$ algebra  of the single-particle observables. It is shown that  a range of the  measurement strength exists, where on one hand the dynamics of the target observables is negligibly affected by the measurement but on  the other hand the measurement is strong enough to localize solutions of the sNLSE in the associated phase-space. 

The localization leads to a drastic reduction of the computational complexity of the sNLSE compared to the original Schr\"odinger equation. The algorithm employs a time-dependent basis of the spin-coherent states. Asymptotically the number of states in the  basis  scales logarithmically with the Hilbert space dimension $n=N+1$ for the simulation on the time-interval of order $\ln n$. Therefore, the memory resources and the cost of an elementary time-step of the simulation are reduced by the factor $\frac{n}{\ln n}$ compared to conventional sparse-matrix algorithms for solving the corresponding Schr\"odinger Equation. The number of the time-steps per unit time for the convergent simulation on the fixed time interval scales as  $n^{1/2}$, i.e., is smaller by the factor $n^{1/2}$ than in conventional methods.
This leads to the overall scaling of $O(\sqrt{n} \ln^3 n)$ of the computational resources for solving the sNLSE.  The estimate of $O(n^{1/2})$ scaling of the number of time-steps  is rooted in the present  linear implementation of the algorithm. It is  possible that the more flexible nonlinear implementation will lead to a  better scaling. 
Solution of the sNLSE corresponds to a single realization of the surrogate open system dynamics. To obtain  a reasonable accuracy of the simulation solutions of the sNLSE are averaged over $\sim 10^3$  realizations.  Though in practice the averaging increases substantially the computational cost of the simulation, the number of  realizations is independent on $n$ and does not contribute to the asymptotic scaling.
 An important additional feature of the algorithm is that  realizations of the open system dynamics can be simulated in parallel, which is  advantageous from the computational standpoint. 

Simulating dynamics on the time-interval of order $\ln n$ is necessary to observe the generalized purity loss, corresponding to depletion of the condensate, which is not seen in a mean-field approximation. Therefore, the $\ln n$ factor cannot be removed from the overall scaling of the algorithm and an \textit{efficient} simulation of the depletion of the condensate is impossible in the sense of our definition. It is conjectured that generally a logarithmic factor is present in the scaling of the computational resources with the Hilbert space dimension in a simulation of quantum many-body effects, i.e., effects that cannot be accounted for by a mean-field dynamics. This is because generically the mean-field approximation is broken when the generalized coherent state with respect to the SGA approaches the hyperbolic fixed point of the associated mean-field dynamics. The  breakdown sets in when  the distance to the fixed point becomes of the order of the total uncertainty of the generalized coherent state. The corresponding time-scale is of the order of the logarithm of the total uncertainty, which is of the order of the maximal weight of the Hilbert space representation of the SGA, increasing with the Hilbert space dimension. 

To summarize, the  algorithm described in the present work reduces the memory resources, the cost of an elementary time-step of the simulation and  the CPU time resources compared to conventional sparse-matrix approaches. As a consequence, asymptotically, it outperforms conventional sparse-matrix computations by the factor  $\frac{n^{3/2}}{\ln n}$. It is conjectured that implementation of the algorithm to simulate many-body dynamics with other SGA will prove advantageous as well.
\begin{acknowledgments}
We thank  Uri Heinemann for his invaluable help in programming and Shimshon Kallush for sharing  results of simulation using his sparse-matrix algorithm.
Work supported by DIP and the Israel Science Foundation (ISF).
The Fritz Haber Center is supported
by the Minerva Gesellschaft f\"{u}r die Forschung GmbH M\"{u}nchen, Germany.

\end{acknowledgments}

 \end{document}